\documentclass[%
 preprint,
superscriptaddress,
 amsmath,amssymb,
 aps,
]{revtex4-1}
\usepackage{lipsum}
\usepackage{graphicx}
\usepackage{dcolumn}
\usepackage{bm}
\usepackage{xcolor}
\usepackage{verbatim}

\usepackage{hyperref}

\usepackage{epstopdf}
\AppendGraphicsExtensions{.tif}
\usepackage{color,soul}

\begin{document}


\title{The role of interfacial interactions and oxygen vacancies in tuning magnetic anisotropy in LaCrO$_{3}$/LaMnO$_{3}$ heterostructures}
\author{Xuanyi Zhang}
 \email{xuanyi.zhang@duke.edu}
 \affiliation{Department of Physics, Duke University, Durham, NC 27517, USA}

 \author{Athby Al-Tawhid}
 \affiliation{Department of Physics, North Carolina State University, Raleigh, NC 27695, USA}

 \author{Padraic Schafer}
 \affiliation{Advanced Light Source, Lawrence Berkeley National Laboratory, Berkeley, California 94720, USA}

  \author{Zhan Zhang}
\affiliation{Advanced Photon Source, Lemont, IL 76019, USA}

\author{Divine P. Kumah}
\email{divine.kumah@duke.edu}
  \affiliation{Department of Physics, Duke University, Durham, NC 27517, USA}
  \affiliation{Department of Physics, North Carolina State University, Raleigh, NC 27695, USA}
 

\date{\today}

\begin{abstract}
The interplay of lattice, electronic, and spin degrees of freedom at epitaxial complex oxide interfaces provides a route to tune their magnetic ground states. Unraveling the competing contributions is critical for tuning their functional properties. We investigate the relationship between magnetic ordering and magnetic anisotropy and the lattice symmetry, oxygen content, and film thickness in compressively strained LaMnO$_3$/LaCrO$_3$ superlattices. Mn-O-Cr antiferromagnetic superexchange interactions across the heterointerface resulting in a net ferrimagnetic magnetic structure. Bulk magnetometry measurements reveal isotropic in-plane magnetism for as-grown oxygen-deficient thinner thin samples due to equal fractions of orthorhombic a+a-c-, and a-a+c- twin domains. As the superlattice thickness is increased, in-plane magnetic anisotropy emerges as the fraction of the a+a-c- domain increases. On annealing in oxygen, the suppression of oxygen vacancies results in a contraction of the lattice volume, and an orthorhombic to rhombohedral transition leads to isotropic magnetism independent of the film thickness. The complex interactions are investigated using high-resolution synchrotron diffraction and X-ray absorption spectroscopy. These results highlight the role of the evolution of structural domains with film thickness, interfacial spin interactions, and oxygen-vacancy-induced structural phase transitions in tuning the magnetic properties of complex oxide heterostructures.  

\end{abstract}

\maketitle

\section{Introduction}
Tunable magnetism in low-dimensional systems has important implications for the design of high-performance spintronic logic and memory devices.\cite{wang2018ferroelectrically, zheng2022high, huang2021novel} While achieving a strong perpendicular magnetic anisotropy in complex oxide heterostructures has garnered significant interest, material systems with a strong in-plane magnetic anisotropy have distinct applications in spintronics devices such as three-terminal spin-orbit torque (SOT) devices\cite{fukami_spinorbit_2016,luo_simultaneous_2017}, enabling high-density skyrmion transfer in magnetic wires\cite{moon_existence_2019}, and magnetoresistive random-access memory (MRAM) device based binary stochastic neurons (BSNs)\cite{hassan_low-barrier_2019}. The tunability of in-plane magnetic anisotropy can significantly impact the performance and functionality of these devices.

At epitaxial 3\textit{d} transition metal oxide perovskite interfaces, the competing effects of chemical, structural, spin, and electronic interactions provide knobs for tuning magnetic ordering and magnetic anisotropy.\cite{Hellman2017Interface-inducedMagnetism,10.1063/1.123634,yi2016atomic,guo2018control, Bhattacharya2014MagneticHeterostructures, vaz2009magnetic, feng2023tuning, santos2011delta, li2017impact} The magnetic ground state depends on the exchange interactions between the transition metal \textit{3d} orbitals and the ligand Oxygen \textit{2p} orbitals which are inherently linked with the electronic occupation of the transition metal d orbitals and rotations and deformations of the oxygen octahedra (OO) surrounding the transition metal ion.\cite{PhysRevLett.111.057202,PhysRevB.93.224401} Additionally, magnetocrystalline anisotropy arising from spin-orbit interactions depends on strain-induced deformations of the oxygen octahedra and differences in the magnitude and phase of rotations along the principal crystalline axes.\cite{PhysRevLett.119.077201,singh2018spin} Control of oxygen octahedral distortions at interfaces via epitaxial strain and symmetry-mismatched buffer layers has been shown in recent years to induce magnetic phase transitions. For example, the introduction of a cubic SrTiO$_3$ (STO) buffer between ferromagnetic rhombohedral LaSrMnO$_3$ (LSMO) and orthorhombic NdGaO$_3$ substrates modifies the magnitude of the oxygen octahedral (OO) rotations in the LSMO leading to a thickness-dependent realignment of the magnetic easy axis.\cite{liao2016controlled} A similar tuning has been observed at SrRuO$_3$/CaSrTiO$_3$ interfaces\cite{kan_tuning_2016}, LaVO$_3$/SrVO$_3$ superlattice\cite{LUDERS20141354} and LSMO/SrIrO$_3$ superlattice\cite{PhysRevApplied.15.024001}.

While strain and symmetry-mismatch at interfaces are effective routes to tune magnetocrystalline anisotropy, decoupling the competing contributions of stoichiometry and lattice symmetry to the magnetic anisotropy remains a challenge. In the bulk canonical system LaMnO$_3$ (LMO), the stoichiometric compound is orthorhombic (\textit{Pbnm}) with out-of-phase OO tilts along the pseudocubic (100)$_{c}$ and (001)$_{c}$ axes, and in phase OO rotations along the (001)$_{c}$ axes. (The subscript \textit{c}, refers to the pseudocubic lattice). Stoichiometric bulk LMO is an A-type antiferromagnet with aligned spins along the [001]$_{c}$ and [100]$_{c}$ directions and anti-parallel spins along the [100]$_{c}$ directions. The magnetic ordering in LMO has been studied extensively theoretically and experimentally.\cite{millis1997orbital, lee2013strong, mellan2015importance} Ferromagnetism has been observed in LMO thin films and has been attributed to epitaxial strain, interfacial charge transfer, and doping induced by oxygen vacancies.\cite{Solovyev1996Crucial3, Chen2017ElectronHeterostructures, roqueta2015strain, Gibert2012ExchangeSuperlattices}  Magnetic ordering in LMO thin films on STO can be tuned from a ferromagnetic state with a Curie temperature of about 200 K to an antiferromagnetic state by varying the growth oxygen partial pressures and introducing capping layers. \cite{roqueta2015strain, Wu2017Interface-inducedFilms,Garcia-Barriocanal2010SpinInterfaces, Aruta2006PreparationDeposition}

\section{Results and Discussion}
In the current work, we demonstrate the tunable in-plane magnetic anisotropy in compressively-strained LaCrO$_3$(LCO)/LaMnO$_3$(LMO) (LCO/LMO)$_N$ superlattices where N is the number of bilayer repeats. Stoichiometric LMO and LCO are A-type and G-type antiferromagnetically ordered in bulk, respectively. We find at the LCO/LMO interface, antiferromagnetic Cr-O-Mn superexchange leads to ferromagnetically ordered LMO layers, evidenced by X-ray magnetic circular dichroism (XMCD) measurements. The angular-dependent magnetism is investigated by SQUID magnetometry and shows changes in the in-plane magnetic anisotropy which are correlated with the oxygen stoichiometry, film thickness and OO rotation symmetry. A strong uniaxial in-plane magnetic anisotropy is observed for oxygen-deficient thicker films (N=20) with the magnetic hard axis along the [100] direction while magnetism in thinner films (N=10) is isotropic. The thickness-dependence of the magnetic anisotropy for the oxygen-deficient films is related to the presence of orthorhombic a+a-c- and a-a+c- domains whose relative fractions are dependent on the film thickness. Annealing in oxygen leads to isotropic magnetism independent of the film thickness related to an orthorhombic-rhombohedral structural transition and a reduction of the c/a ratio. Our finding leads to a better understanding of the microscopic origin of the magnetic anisotropy in (LCO/LMO) superlattices and demonstrates a pathway to engineering the magnetic anisotropy with oxygen vacancy in transition-metal oxide heterostructures.

\begin{figure*}[t]
\centering
\includegraphics[width=6.5in]
{ 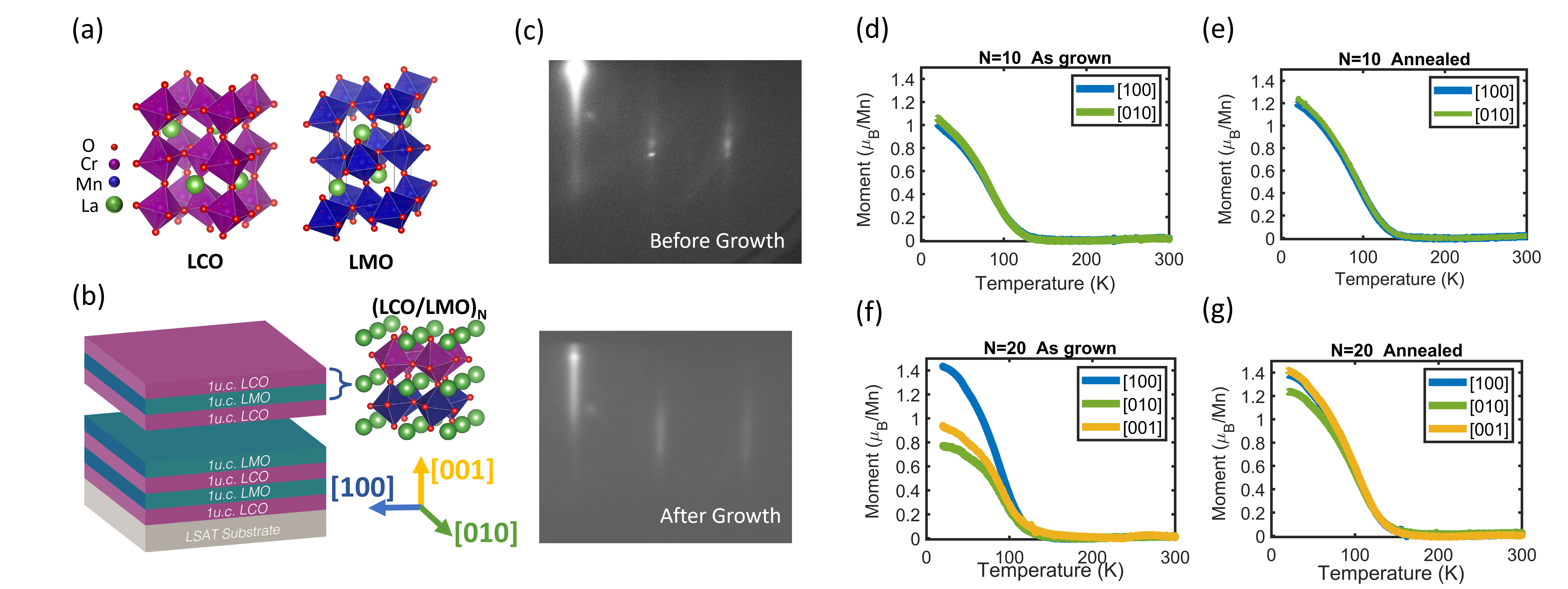}
\caption{(a)Schematic diagram of bulk LaCrO$_3$ (LCO) and bulk LaMnO$_3$(LMO (b) Schematic diagram of LCO/LMO bilayer superlattice grown on (001)-oriented LSAT substrates by MBE. (c) RHEED pattern of initial LSAT before growth and film after growth. Magnetization vs temperature for (LCO/LMO)$_{N}$ films on LSAT substrate measured at 4000 Oe on warming after field cooling in a 4000 Oe magnetic field for as-grown (d)  N=10 (f) N=20 and oxygen-annealed  (e)N=10 and (g) N=20.}
\label{fig:RHEED}
\end{figure*}

Superlattices comprising of N-repeats of (1 unit cell LCO/1 unit cell LMO) bilayers were grown on (001)-oriented (La$_{0.18}$Sr$_{0.82}$)(Al$_{0.59}$Ta$_{0.41)}$O$_3$ (LSAT) substrates by molecular beam epitaxy (MBE). Representative reflection high-energy electron diffraction (RHEED) images of the LSAT substrate and (LCO/LMO) film are shown in Figure \ref{fig:RHEED} (c). The observed sharp RHEED streaks are indicative of a fully strained epitaxial film. The magnetic and structural properties of the films were investigated for the as-grown samples and after annealing in 1 atm high purity oxygen at 650 $^oC$ for 6 hours to minimize oxygen vacancies.

The magnetic properties of the films were investigated with SQUID magnetometry before and after the post-growth anneal. Figure \ref{fig:RHEED} (d) - (g) shows the measured temperature-dependent magnetization along the principal pseudo-cubic axes for the (LMO/LCO)$_{10}$ and (LMO/LCO)$_{20}$ samples. A paramagnetic to ferromagnetic transition is observed at 125 K for both samples. The magnetization for the (LMO/LCO)$_{10}$ sample is isotropic along the in-plane [100]$_{c}$ and [010]$_{c}$ directions before (Figure 1(d)) and after (Figure 1(e)) the oxygen anneal. There is a slight increase in the saturation magnetization from 1 $\mu$ B/Mn for the as-grown (LMO/LCO)$_{10}$ sample along the principal pseudo-cubic axes to 1.2 $\mu$B/Mn after annealing. On the other hand, for the as-grown (LMO/LCO)$_{20}$ sample, a strong in-plane uniaxial magnetic anisotropy is observed with the magnetic easy axis along the [100]$_c$ direction (Figure 1 (f)) and annealing leads to uniform magnetism (Figure 1(g)). 

The saturation magnetization values are less than 3.5 $\mu_B$/Mn reported for ferromagnetic LMO\cite{roqueta2015strain} films suggesting that the net magnetization may be due to a canted ferrromagnetic state, or anti-parallel spin contributions from the LCO layers.\cite{koohfar2021interface, Koohfar2019Confinement} The LCO contribution to the net magnetic moments is confirmed by element-specific XMCD measurements.

\begin{figure*}[ht]
\centering
\includegraphics[width=6in]
{ 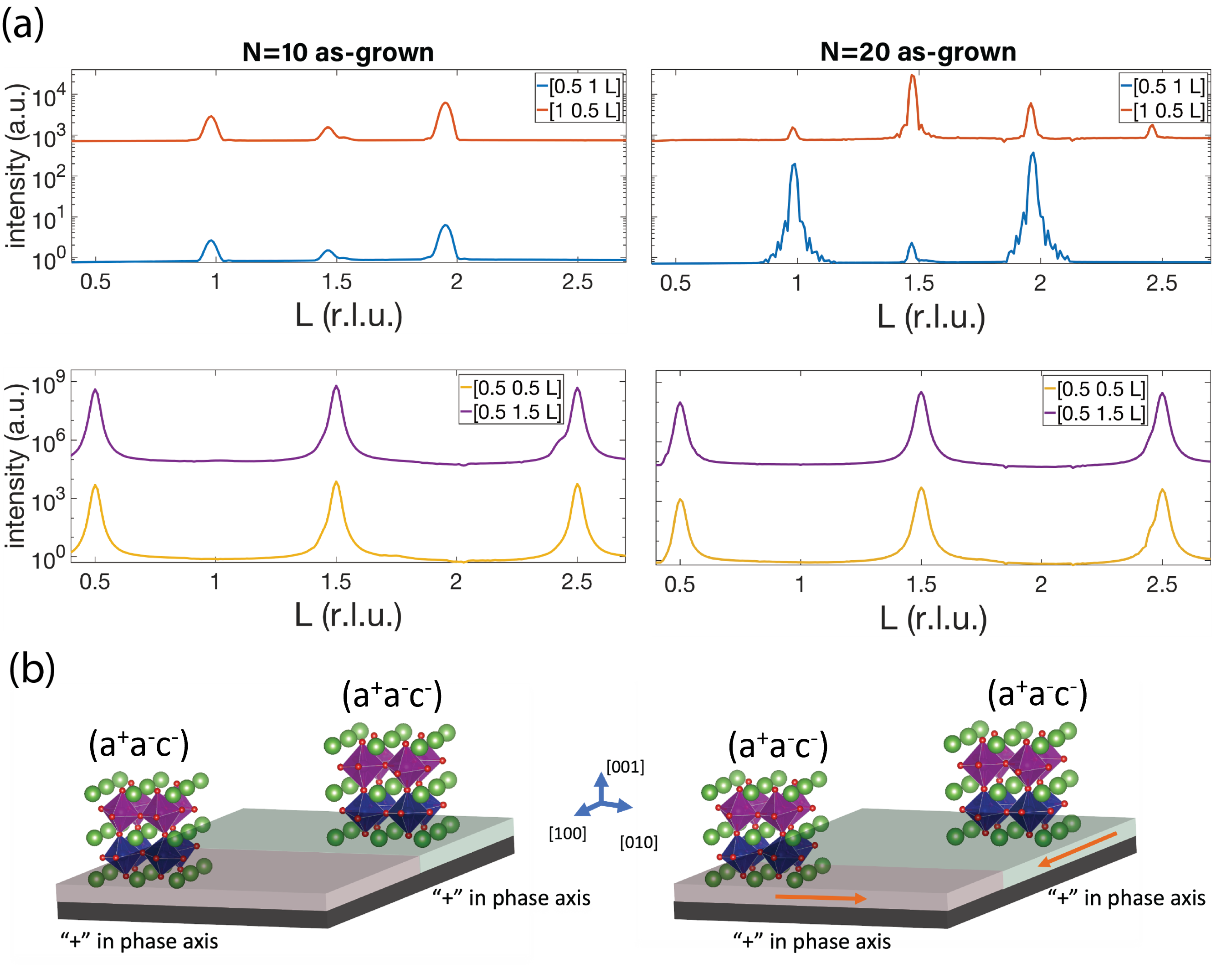}
\caption{(a) Measured half-order X-ray diffraction intensities of as-grown (LMO/LCO)$_{10}$ and (LMO/LCO)$_{20}$ along the substrate-defined reciprocal lattice vector L (1 r.l.u. = 1/0.3865 nm$^{-1})$(b) Schematic diagram of the (a+a-c-) and (a-a+c-) domain volume ratio. The left diagram shows 1:1 domain volume ratio in the (LMO/LCO)$_{10}$ sample and the right diagram shows 6:1 domain volume ratio in the (LMO/LCO)$_{20}$ sample.}
\label{fig:domains}
\end{figure*}

 Due to the strong coupling of the structural and spin degrees of freedom in the system, the structural properties of the samples were investigated with high-resolution synchrotron-based X-ray diffraction measurements. The MnO$_{6}$ and CrO$_{6}$ octahedral rotation configuration of the samples were determined by the measured half-order reflections. Figure \ref{fig:domains} (a) shows a comparison of the [0.5 1 L] and [1 0.5 L] rods for the as-grown (LMO/LCO)$_{10}$ and (LMO/LCO)$_{20}$ samples. Finite thickness oscillations are observed along the half-order rods indicative of the excellent crystallinity of the samples. The peaks along these rods are indicative of in-phase octahedral rotations in the in-plane direction. In-phase octahedral rotations about the [100] axis, denoted by a+ in Glazer notation, give rise to peaks of type (\textit{o/2,e/2,e/2})$_{c}$ and (\textit{e/2,o/2,o/2})$_{c}$  where \textit{e} and \textit{o} are even and odd integers, respectively.\cite{Woodward:dr5005} La ions are displaced along directions orthogonal to the in-phase rotation axis.\cite{zhang2022thickness} Conversely, peaks of type (\textit{o/2,e/2,o/2})$_{c}$ and (\textit{e/2,o/2,e/2})$_{c}$ are indicative of in-phase rotations about the pseudocubic [010]$_c$ direction.  Peaks of both types are present for the (LMO/LCO)$_{10}$ and the (LMO/LCO)$_{20}$ samples as shown in Figure \ref{fig:domains}(a) indicative of the presence of domains of octahedral rotation patterns of type (a+a-c-) and (a-a+c-). The rhombohedral LSAT substrate has out-of-phase (a-a-a-) rotations giving rise to \textit{(o/2, o/2, o,2)} peaks which overlap with the film peaks as shown in the lower panels of Figure \ref{fig:domains}(a).

The relative fractions of domains are quantitatively determined by the intensities of the corresponding half-ordered peaks. The intensities along the [0.5 1 L] and [1 0.5 L] rods for the as-grown (LMO/LCO)$_{10}$ sample (Figure \ref{fig:domains}) are approximately equal, indicating equal fractions of (a+a-c-) and (a-a+c-) domains. For the as-grown (LMO/LCO)$_{20}$ sample, the intensities along both rods are not equal and the ratio of the (a+a-c-) and (a-a+c-) fractions is determined to be 6:1 as illustrated in Figure \ref{fig:domains}(b). The tendency to form a single domain as the thickness increases is consistent with results for other orthorhombic thin films including SrRuO$_3$.\cite{zhang2022thickness} 

The magnitude of the OO rotations for the as-grown samples at room temperature are obtained from fits to the half-order rods using the GenX genetic fitting algorithm.\cite{bjorck2007genx} The structural fit results are summarized in Table \ref{tab:fitresults}. The OO rotation angle ($\gamma$) about the [001] axis is larger than the tilt angles ($\alpha$,$\beta$) about the in-plane axes due to the compressive strain exerted on the LCO and LMO layers by the LSAT substrate. The rotation angles are comparable for the a+ and b- directions, thus, we postulate that the magnetic anisotropy along the in-plane axes is related to the sense of the rotations and not the magnitude of the rotations.

 \begin{figure*}[ht]
\centering
\includegraphics[width=6in]
{ 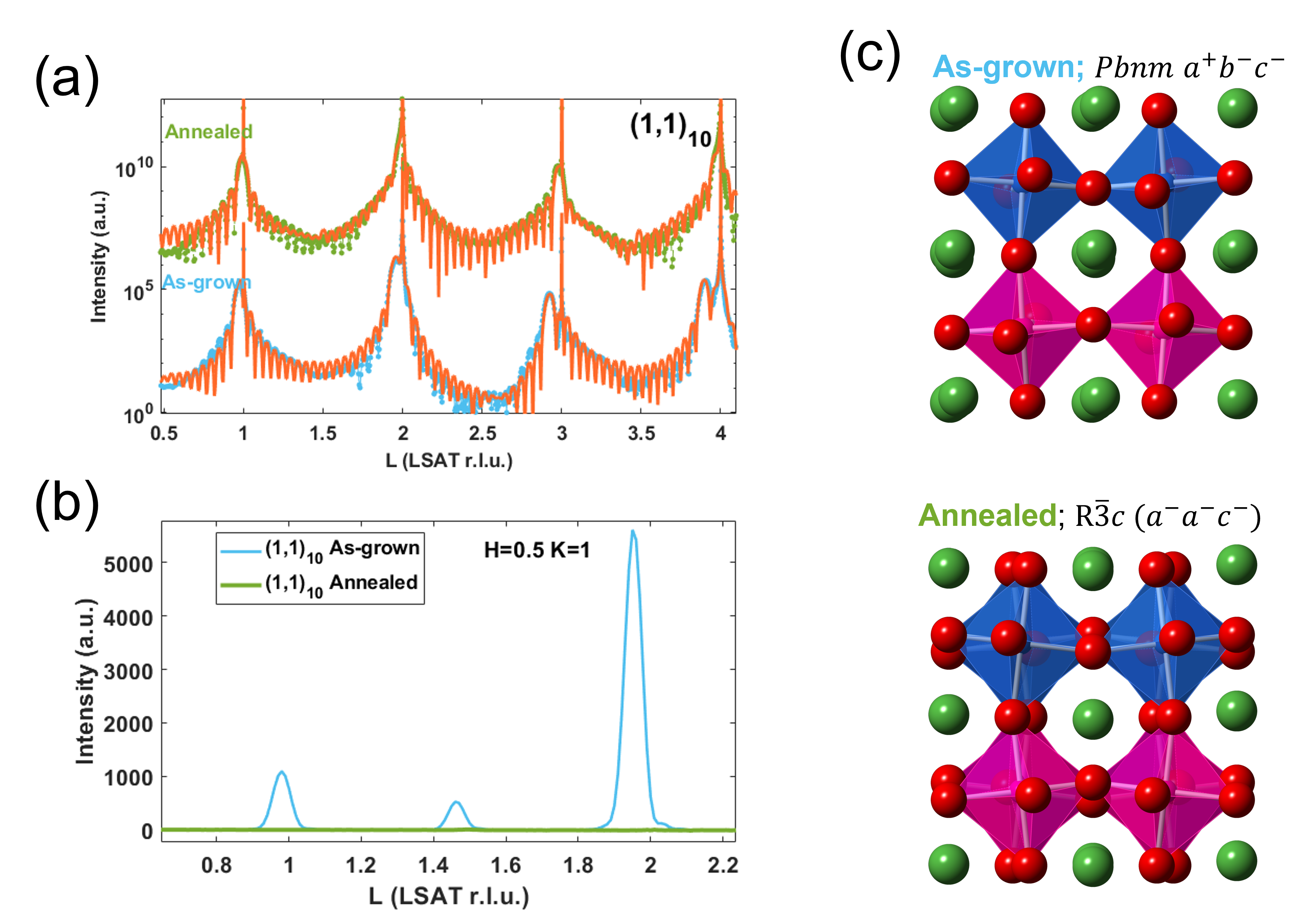 }
\caption{(a) Comparison of (00L) crystal truncation rods for (LCO/LMO)$_{10}$ superlattices before (as-grown) and after an anneal in $O_2$. The circles are the measured intensities, and the solid lines represent fits to the data. (b) Comparison of half-order (0.5 1 L) reflections before and after oxygen annealing related to La displacements and in-phase octahedral tilts for the (LCO/LMO)$_{10}$ superlattice. (c) Schematic diagrams of the resulting crystal structures projected along the [100]$_c$ direction.}
\label{fig:lattice_constant2}
\end{figure*}

To further understand the effect of annealing on the structural properties of the superlattices, we compare the (00L) CTR and the half-order peaks for the (LMO/LCO)$_{10}$ sample in Figure \ref{fig:lattice_constant2} (a) and (b). The LMO (LCO) out-of-plane lattice parameter, determined through fitting the CTRs, contracts from 4.048{\AA}(3.902{\AA}) to 3.895{\AA} (3.899{\AA}) following the anneal for the (LMO/LCO)$_{10}$. With the in-plane lattice constant a = 3.865{\AA} for the coherently strained film on LSAT substrate, the LMO c/a ratio reduces from 1.047 to 1.008 and the unit cell volume decreases from 60.6{\AA}$^3$ to 58.3{\AA}$^3$. The reduction of lattice parameters is also observed in the (LMO/LCO)$_{20}$. The contraction of the out-of-plane lattice constant is indicative of a decreased concentration of oxygen vacancies.\cite{ritter1997influence}

In addition to the reduction in the lattice volume, annealing leads to a suppression of the orthorhombic distortion as evidenced by a suppression of the  $(o,e,L)$ peaks for the annealed samples as shown for the (\textit{0.5 1 L}) rod in Figure \ref{fig:lattice_constant2} (b). The octahedral rotation configuration of oxygen-annealed superlattices switches from (a+a-c-) in the orthorhombic \textit{Pbnm} phase (Figure \ref{fig:lattice_constant2}(c))  to (a-a-c-) in the rhombohedral phase (Figure \ref{fig:lattice_constant2}(d)).

\begin{table}[]
    \centering
    \begin{tabular}{|c|c|c|c|c|}
     \hline
         Parameters                   &(LMO/LCO)$_{10}$                   &	(LMO/LCO)$_{20}$  \\
         \hline 
          c$_{avg}$ (\AA{})                           &3.975              &	3.935\\ \hline

          LCO $\alpha_{rot}$ (in-phase)           &6.47$^o$                  & 6.39$^o$     \\
          LMO $\alpha_{rot}$ (in-phase)           &4.25$^o$                  & 4.46$^o$     \\
          LCO $\beta_{rot}$ (out-of-phase)        &6.58$^o$                  & 6.02$^o$     \\
          LMO $\beta_{rot}$ (out-of-phase)        &4.33$^o$                  & 4.12$^o$     \\
          LCO $\gamma_{rot}$ (out-of-phase)       &14.94$^o$                  & 14.90$^o$     \\
          LMO $\gamma_{rot}$ (out-of-phase)       &12.61$^o$                  & 12.52$^o$     \\
          \hline
          $Mn-O-Mn$(100)           &153.36$^o$                       & 153.66$^o$     \\
          $Mn-O-Mn$(010)           &153.30$^o$                       & 153.86$^o$     \\
          $Cr-O-Cr$(100)           &147.66$^o$                       & 147.58$^o$     \\
          $Cr-O-Cr$(010)           &147.57$^o$                       & 147.84$^o$     \\
      
            \hline
       
    \end{tabular}
    \caption{Comparison of structural parameters of as-grown LCO/LMO superlattices. Error bars of the OOR angles are within $\pm 2^o $, corresponding to the parameter range for a $\pm 5\% $ variation in the figure of merit of the fitted model.}
    \label{tab:fitresults}
\end{table}

\begin{figure*}[ht]
\includegraphics[width=5in]{ 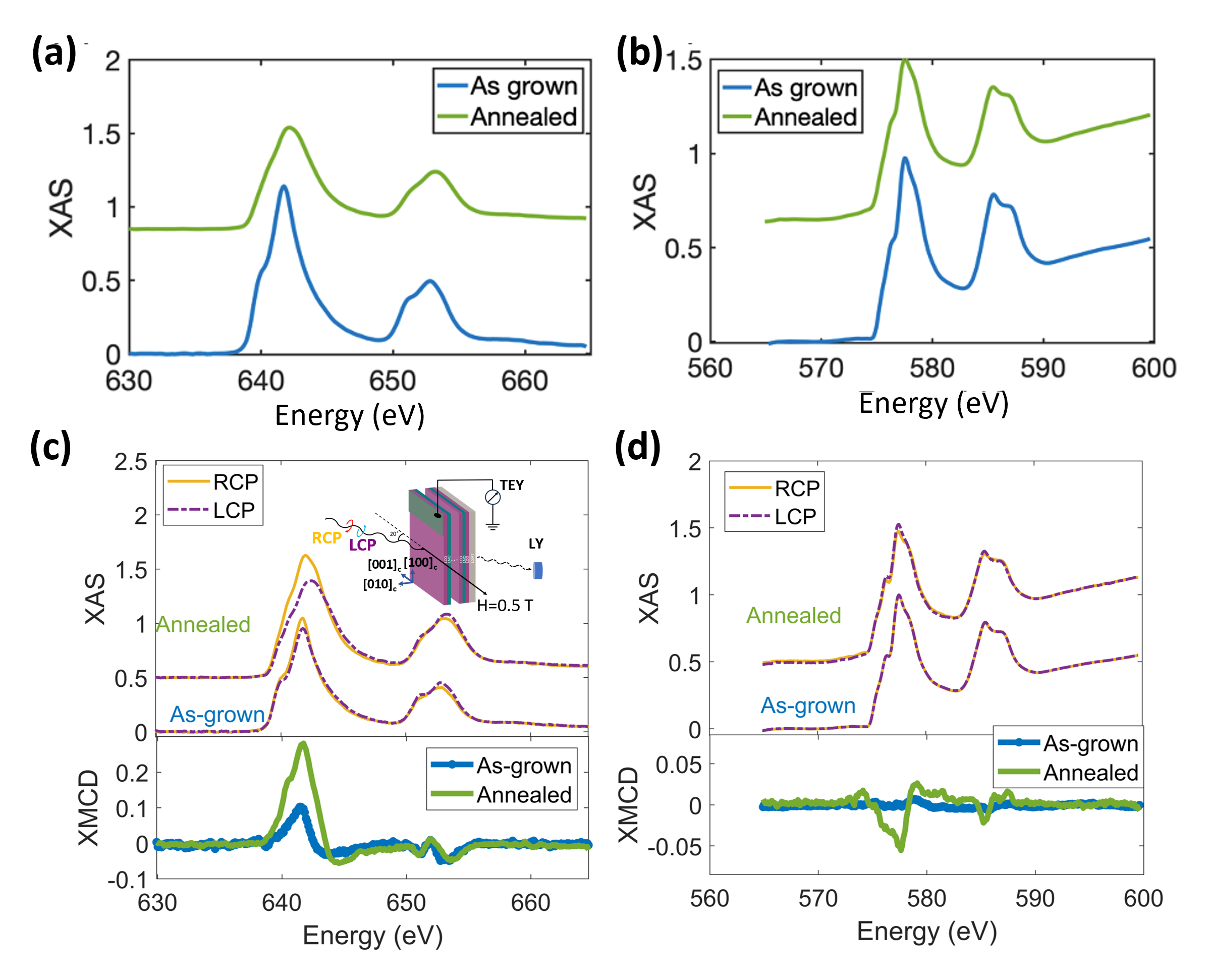} 
\caption{ Comparison of XAS measured at 300 K at the (a) Mn L-edge and (b) Cr L edge for as-grown and annealed (LMO/LCO)$_{20}$. XMCD measurements at the (c) Mn L-edge and (d) Cr L-edge. The XMCD is determined from the difference in the XAS measured for right and left circularly polarized incident light.}
\label{fig:XMCD}
\end{figure*}

The in-plane magnetic anisotropy observed for both thicknesses after annealing is correlated with (1) the suppression of oxygen vacancies (2) a reduction in the unit-cell volume and (3) a suppression of the orthorhombic distortion. To decouple the contributions of these factors to the magnetic anisotropy, element-specific X-ray absorption (XAS), XMCD, and X-ray linear dichroism (XLD) measurements are compared as a function of thickness and oxygen stoichiometry. 

XAS measurements are performed at the Mn and Cr L edges at 300 K and compared in Figure \ref{fig:XMCD}(a) and \ref{fig:XMCD}(b), respectively, for the as-grown and annealed (LCO/LMO)$_{20}$ sample. For the as-grown samples, the Mn spectra indicates a significant fraction of Mn$^{2+}$ evidenced by the shoulder at 640 eV, to compensate for the oxygen vacancies present in the films. On annealing the samples in oxygen, the Mn$^{2+}$ converts to Mn$^{3+}$ as evidenced by the reduction in the shoulder intensity at 640 eV. The Cr spectra are not affected by the annealing process indicating that the oxygen vacancies are primarily located in the LaMnO$_{3-\delta}$ layers.

Magnetic ordering in the LMO and LCO sublattices are probed by element-specific XMCD measurements with an 0.5 T magnetic field applied along the magnetic hard axis ([010]$_c$) identified by SQUID measurements and XRD. The XMCD spectra is determined by the difference in X-ray absorption for parallel and antiparallel alignment of the photon helicity and magnetic moments of Mn and Cr. The XMCD measurements are performed in luminescence yield below the paramagnetic-ferrimagnetic transition (Tc=125 K) at 80 K. The XMCD results at the Mn and Cr L-edges are compared in Figure \ref{fig:XMCD}(c) and (d), respectively, for the (LCO/LMO)$_{20}$ sample before and after the post-growth anneal. The Cr and Mn XMCD signals along the hard axis increase after annealing. The Mn orbital ($m_o$) and spin ($m_s$) and total($m=m_o+m_s$) magnetic moments at 80 K are calculated using the XMCD sum rules assuming 4 holes on the Mn$^3+$ site. The measured SQUID and XMCD moments are compared in Table \ref{tab:magneticmoment}. For the (LCO/LMO)$_{10}$, the measured moments at 80 K are $m=$0.63 $\mu_B$ ($m_{o}/m_{s}=0.12$) and $m=$0.77 $\mu_B$ ($m_{o}/m_{s}=0.10$) for the as-grown and annealed sample. The slight increase in the moment for the N=10 sample on annealing is consistent with the SQUID results. 
For the (LCO/LMO)$_{20}$ sample, the measured moments along the [010]$_c$ are  $m=$0.64 $\mu_B$ ( $m_{o}/m_{s}=-0.15$) and $m=$1.58 $\mu_B$ ($m_{o}/m_{s}=0.15$) for the as-grown and annealed samples, respectively. Here, the increase by a factor of $\sim$2 is consistent with the change in the magnetic anisotropy for the N=20 sample. The orbital moments are anti-aligned with the spin moments for the as-grown sample and aligned for the annealed N= 20 sample. The significant orbital moment is evidence for spin-orbit coupling (SOC) described by the Hamiltonian term $H_{SO}=\lambda L \cdot S$, where $\lambda$ is the SOC strength, $L$ is the spin angular momentum operator and $s$ is the orbital angular momentum operator.\cite{PhysRevB.92.054106, yi2016atomic, Garcia-Barriocanal2010SpinInterfaces} The magnetic anisotropy suggests that the spin-orbit coupling is anisotropic along the [100] and [010] directions in the as-grown oxygen-deficient $Pbnm$ phase.

\begin{table}[]
    \centering
    \begin{tabular}{|c|c|c|c|c|}
     \hline
         (LCO/LMO)$_{10}$  \newline   &As-grown  $(\mu_B/Mn)$                  & Annealed $(\mu_B/Mn)$\\
         \hline \hline
          SQUID  [100]                   &0.52                   &0.71\\ 
           SQUID  [010]                     &0.51                   &0.69\\ 
          XMCD  $m_s$                    &0.56                   &0.69\\
          XMCD  $m_o$                    &0.07                   &0.07\\ \hline
          \hline
         (LCO/LMO)$_{20}$                  &As-grown                   &Annealed\\
         \hline \hline
          SQUID  [100]                   &0.79                   &0.90\\ 
          SQUID  [010]                    &0.49                   &0.83\\ 
          XMCD $m_s$                   &0.68                   &1.37\\
          XMCD $m_o$                   &$-0.04$                   &0.15\\ \hline
    \end{tabular}
    \caption{Comparison of SQUID and XMCD properties of LCO/LMO superlattices at 80 K, the spin and orbital magnetic moment measured by XMCD is along the magnetic hard axis}
    \label{tab:magneticmoment}
\end{table}

On annealing, a Cr XMCD signal emerges with a sign opposite to that of the Mn XMCD signal due to the antiferromagnetic exchange between the Mn$^{3+}$ and the Cr$^{3+}$ orbitals across the LCO/LMO interface. The antiferromagnetic Cr-O-Mn exchange is consistent with predictions based on the Goodenough-Kanamori rules for the half-filed Mn eg orbital and the unoccupied Cr eg-orbital. The calculated Cr moment is -0.18 $\mu_B$ suggestive of canting of the Cr spins.

\begin{figure*}[ht]
\includegraphics[width=6in]{ 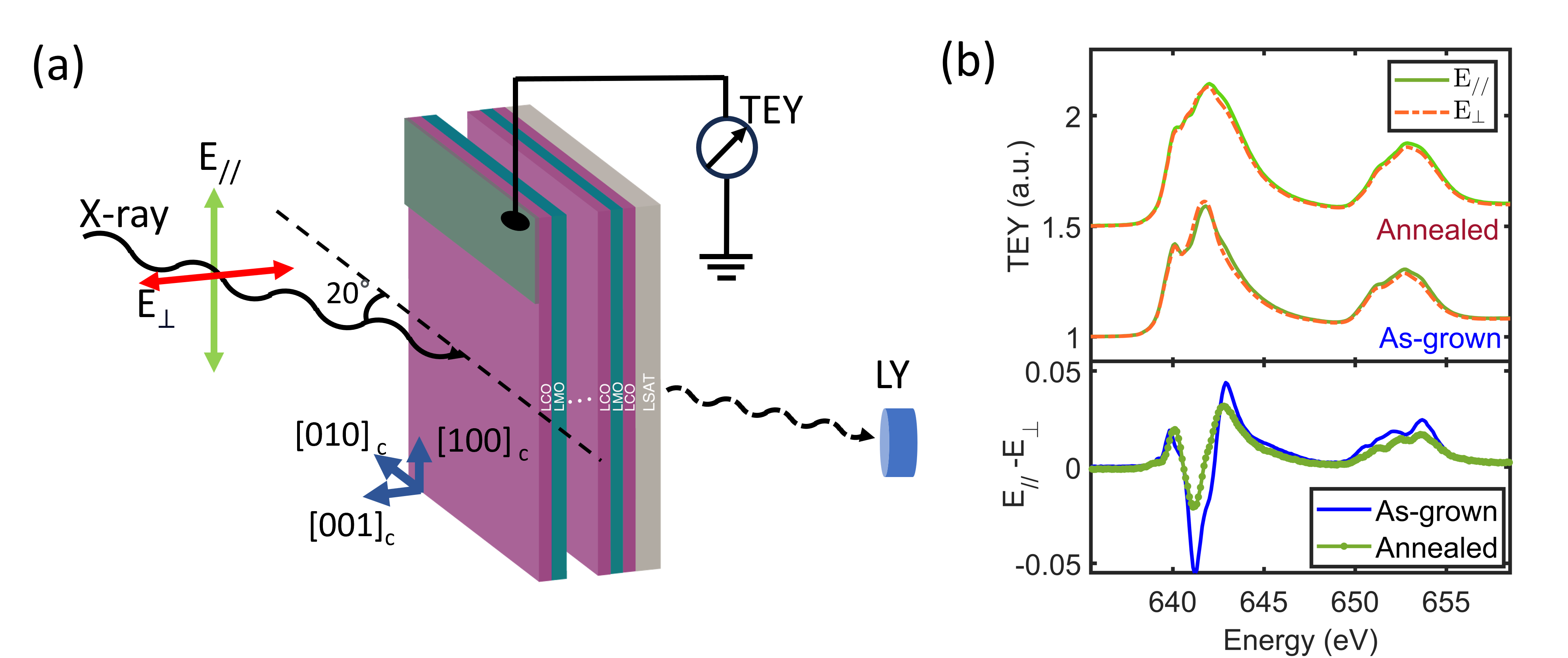} 
\caption{  (a) Schematic diagram of XLD experiment set up. The X-ray absorption is measured in total electron yield (TEY) and luminescence yield (LY) mode. (b) XLD spectra in TEY mode at the Mn L-edge at 300 K for as-grown and oxygen-annealed (LCO/LMO)$_{20}$ with the linear polarization of the incident X-rays along the in-plane [100] direction ($E_{//}$) and the out-of-plane [001] direction ($E_{\perp}$). The differences in the spectra are plotted in the lower panel. }  
\label{fig:XLD}
\end{figure*}

To further understand the impact of the structural transitions induced by the annealing process, XLD measurements at the Mn and Cr L-edges were performed with linearly polarized light where the polarization direction was aligned either normal to the film surface along the [001]$_c$ direction ($E_\perp$) or parallel to the film surface ($E_{//}$) along the [100]$_c$ crystallographic direction. The integrated difference absorption spectra measured for $E_{//}$ and $E_\perp$is proportional to the relative orbital occupations of the in-plane oriented orbitals ($d_{x^2-y^2}$), and the out-of-plane oriented orbitals ($d_{z^2-r^2}$). Figure \ref{fig:XLD}(b) shows XLD for the(LCO/LMO)$_{20}$ sample before and after the post-growth anneal at the Mn L-edge. The integrated intensity is positive, indicative of the breaking of the degeneracy of the $e_g$ orbitals with a preferential electron occupation of the out-of-plane $d_{z^2-r^2}$ orbital. The reduction in the orbital anisotropy by 7\% after annealing is consistent with the decrease in the c/a ratio observed by the XRD measurements.

The emergent relationship between the thickness, magnetic, structural, and electronic properties of the LMO/LCO system is as follows: The Mn and Cr spins are antiferromagnetically coupled across the LCO/LMO interface.\cite{Koohfar2019Confinement, koohfar2021interface} For the as-grown oxygen-deficient samples, the c/a ratios are 1.047 and the films are orthorhombic with (a+a-c-) and (a-a+c-) domains. As the superlattice thickness increases, a single domain emerges leading to a strong in-plane magnetic anisotropy for the as-grown samples. On annealing the samples in oxygen, a reduction of the c/a ratio to 1.008 and a transition from an orthorhombic to rhombohedral structure occurs leading to isotropic magnetism along the in-plane [100]$_c$ and [010]$_c$ directions.

The structural transition observed in the LCO/LMO superlattices is consistent with the symmetry change observed in bulk LMO for different oxygen stoichiometries. Bulk LMO undergoes an orthorhombic $Pbnm$ to rhombohedral $R\bar{3}c$ transition and a reduction in the lattice volume as the oxygen content is increased\cite{ritter1997influence, tiwari2017evolution}. While this transition in bulk is accompanied by an antiferromagnetic to ferromagnetic transition, the ferromagnetic ordering in the strained LMO layers in the LMO/LCO bilayers is a result of the Mn-O-Cr exchange interactions across the LMO-LCO interface. \cite{Koohfar2017StructuralInterfaces, koohfar2021interface} The FM ordering is in contrast to the G-type AF order expected for compressively strained LMO.\cite{nanda2010magnetic, lee2013strong}

\section{Conclusion}
In summary, we have investigated the effect of film thickness and oxygen stoichiometry on the structural and magnetic properties of ferrimagnetic (LCO/LMO)$_N$ superlattices grown on (001)-oriented LSAT substrates. The as-grown samples are oxygen deficient with the oxygen vacancies localized in the LMO layers. The as-grown films are orthorhombic with in-phase octahedral rotations oriented in the plane of the films. For thinner films (N=10), the presence of equal fractions of structural rotational (a+a-c-) and (a-a+c-) domains leads to in-plane magnetic isotropy. On increasing the film thickness (N=20), the films evolve towards a single structural domain leading to enhanced in-plane magnetic anisotropy along the [100] and [010] directions. A post-anneal in oxygen results in a reduction in the concentration of oxygen vacancies and a concomitant orthorhombic-rhombohedral structural transition independent of the film thickness. The in-phase octahedral rotations are suppressed in the rhombohedral phase for the annealed samples resulting in isotropic in-plane magnetization.

The combination of structural, magnetic measurements, and spectroscopy measurements provides a clear picture of the competing roles of oxygen octahedral rotations, oxygen vacancies and strain in modulating the magnetic anisotropy in oxide-based artificial ferrielectrics. These results provide a clear path to tune the magnetic and electronic properties of quantum materials via domain and compositional engineering. The structural and magnetic properties are related to the oxygen stoichiometry, hence, the dynamic modulation can potentially be achieved by gating techniques for spintronic applications.

\section{Experimental Section}

Superlattices comprising of  (1 uc LCO/1 uc LMO)$_{N}$ were grown on (001)$_{c}$-oriented LSAT single crystal substrates by molecular beam epitaxy. Two sets of samples were fabricated with either 10 bilayer repeats (LMO/LCO)$_{10}$ or 20 bilayer repeats, (LMO/LCO)$_{20}$. The La, Mn and Cr high-purity elemental sources were evaporated from Knudsen effusion cells. The fluxes were calibrated prior to growth using a quartz crystal microbalance. The samples were grown at 850$^o$C and in an oxygen partial pressure of 4x$10^{-6}$ Torr. After growth, the samples were cooled to room temperature at the growth oxygen pressure to minimize the formation of oxygen vacancies. The structural and magnetic properties of the films were studied for the as-grown samples and after a post-growth anneal in 1 atm of high-purity oxygen in a tube furnace at 650 $^o$C for 6 hours.

To determine the atomic structures of the superlattices, synchrotron X-ray diffraction measurements were performed at the 33ID beamline at the Advanced Photon Source. Crystal truncation rods (CTRs)\cite{robinson1992surface} and half-order superstructure reflections were measured along directions in reciprocal space defined by the bulk LSAT crystal lattice. The energy of the incident X-ray beam was 16 keV.

The magnetic properties were probed using a Quantum Design superconducting quantum interference device (SQUID). Measurements were carried out in the temperature range of 20 K to 300 K after a 4 kOe field cool.  X-ray absorption spectroscopy (XAS), X-ray linear dichroism (XLD) and X-ray circular magnetic dichroism (XMCD) measurements were performed at the 4.0.2 beamline at the Advanced Light Source. The absorption spectra were measured at the Mn-L and Cr-L edges using TEY and LY measurements at 300 K and 80 K.

\section{Acknowledgements}
This material is based on work supported by the National Science Foundation under Grant No. NSF DMR1751455. This research used resources of the Advanced Light Source, which is a DOE Office of Science User Facility under Contract No. DE-AC02-05CH11231. Use of the Advanced Photon Source was supported by the US Department of Energy, Office of Science, Office of Basic Energy Sciences, under Contract No. DE-AC02-06CH11357. The authors acknowledge use of the SQUID and PPMS facility in the Department of Materials Science and Engineering at North Carolina State University. This work was performed in part at the Analytical Instrumentation Facility (AIF) at North Carolina State University, which is supported by the State of North Carolina and the National Science Foundation (award number ECCS-2025064). This work made use of instrumentation at AIF acquired with support from the National Science Foundation (DMR-1726294). The AIF is a member of the North Carolina Research Triangle Nanotechnology Network (RTNN), a site in the National Nanotechnology Coordinated Infrastructure (NNCI).

\end{document}